\documentclass[twocolumn,english,aps,prl,showpacs]{revtex4}
\usepackage{graphicx}
\usepackage{epsfig}

\begin{document}

\title{Ultrafast Electron Dynamics Theory of Photo-excited Ruthenium Complexes}

\author{Jun Chang, A. J. Fedro, and Michel van Veenendaal}

\affiliation{Department of Physics, Northern Illinois University, De Kalb,
Illinois 60115, USA }

\affiliation{Advanced Photon Source, Argonne National Laboratory, 9700 South
Cass Avenue, Argonne, Illinois 60439, USA}

\date{\today}

\pacs{78.47.J-, 33.50.-j, 82.50.-m  82.37.Vb  82.53.-k}

\begin{abstract}
An explanation is provided for the  ultrafast photo-excited electron dynamics in low-spin Ruthenium (II) organic complexes. The experimentally-observed singlet to triplet decay in the metal-to-ligand charge-transfer (MLCT) states contradicts the expectation that the system should oscillate between the singlet and triplet states in the presence of a large spin-orbit coupling and the absence of a significance change in metal-ligand bond length. This dilemma is solved with a novel quantum decay mechanism that causes a singlet to triplet decay in about 300 femtoseconds.  The decay is mediated by the triplet metal-centered state ($^3$MC) state even though there is no direct coupling between the $^1$MLCT and $^3$MC states. The coupling between the $^3$MLCT and $^3$MC via excited phonon states leads to vibrational cooling that allows the local system to dissipate the excess energy. In the relaxed state, the population of the $^3$MC state is low and the metal-ligand bond length is almost unchanged with respect to the initial photoexcited state, in agreement with experiment.
\end{abstract}
\maketitle
\begin{figure}[t]
\includegraphics[width=0.9\columnwidth]{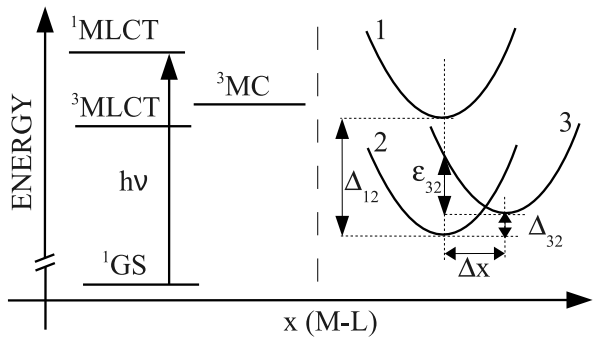}
\caption{Simplified energy level scheme of Ru$^{2+}$-based organic complexes. The initial photon excitation from the singlet ground state ($^1$GS) to the singlet metal-to-ligand charge transfer ($^1$MLCT) state is followed by a relaxation to the long-lived metastable $^3$MLCT state. The triplet metal-centered $^3$MC is located close to the MLCT states.}
\label{level} 
\end{figure}

\textit{Introduction.$-$} Photon-induced nonequilibrium electron dynamics  in transition-metal complexes have drawn intense interest for numerous
applications in diverse research fields such as solar energy conversion, photosynthesis, photodissociation, and photocatalysis \cite{Juris}. The study of the time evolution of the light-driven state is crucial in creating a deeper understanding of these technologically-important phenomena. Often, the photo-induced state undergoes an ultrafast decay into a metastable state from which it relaxes more slowly back into the ground state. A well-studied example is the spin-crossover phenomenon in Fe-complexes, where the spin-orbit coupling causes a cascade from a low-spin state to a high-spin state \cite{Gutlich}. Crucial in the understanding of this process \cite{Veenendaal} is the change of the metal-ligand distance resulting from the transfer of electrons from the  $t_{2g}$ to the $e_{g}$ orbitals that repel the ligands more strongly \cite{Bressler}.
However, a significant bond elongation is not present in all spin-crossover systems.  In ruthenium complexes, such as {[}Ru$^{\text{II}}$(byp)$^{3}$]$^{2+}$, spin flips occur within the $t_{2g}$ orbitals and only a minimal change in metal-ligand bond length is observed \cite{Gawelda}. Due to their wide frequency response, ruthenium compounds have been extensively studied. The ground state is a singlet with six electrons in the $t_{2g}$ orbitals. Since Ru $4d^{6}$ multiplet excitations are dipole forbidden, the optical excitations are dominated by metal-to-ligand charge-transfer (MLCT)  into the $\pi^{*}$ orbitals of the organic ligands, creating a singlet $t_{2g}^5 L$ configuration, where $L$ indicates an electron in the $\pi^*$ ligands. The strong $4d$ spin-orbit coupling then causes spin flips in the $t_{2g}^5$ states leading to triplet MLCT states. Since for the triplet $t_{2g}^5L$ state the coupling to the metal-centered (MC) $t_{2g}^6$ and $t_{2g}^5e_g$ configurations is either forbidden or weak,  it is lower in energy than the antibonding singlet states, see states 1 and 2 in Fig. \ref{level}. The weak coupling has the interesting effect that it tends to localize the photo-excited electron on one of the bipyridine molecules \cite{Damrauer}. The singlet-to-triplet decay is very fast. For example, in {[}Ru$^{\text{II}}$(byp)$^{3}$]$^{2+}$,  the relaxation time is 300 femtoseconds (fs)  \cite{Damrauer,Yeh}. The resulting triplet state is long-lived with a lifetime on the order of hundreds of ns \cite{Creutz,Saes}. 

Despite decades of research, the mechanism for the ultrafast decay from the singlet to the triplet MLCT states is still not understood. The radiationless ultrafast decay is generally accompanied by a pronounced change of the metal-ligand bond lengths \cite{Veenendaal}. However, one does not expect a significant change in the Ru-ligand distance for different $t_{2g}^n$ configurations. This has been confirmed experimentally where for the $^3$MLCT a bond contraction of only $\mathopen{\sim} 0.03$ Angstrom~(\AA)~\cite{Gawelda} with respect to the ground state is observed. As a result, the Huang-Rhys factor (proportional to the square of the bond length change), and hence the decay, between MLCT singlet and triplet is expected to be negligibly small. Therefore, the system should oscillate between the singlet and triplet states via the spin-orbit interaction. Furthermore, it was already noticed \cite{Durham} that a discussion in terms of singlet and triplet states is probably erroneous due to the large spin-orbit coupling with respect to the singlet-triplet energy gap. However, this picture is completely contrary to the standard description of ultra-fast decay between spin-labeled states in Ru-complexes \cite{Damrauer,Yeh} and, in  almost all literature,  a discussion of the mechanism underlying the singlet-to-triplet MLCT decay is generally avoided.

In this Letter, we reconcile these apparent contradictions and present a coherent picture of the singlet-to-triplet decay in the presence of a strong $4d$ spin-orbit-coupling. Crucial in the mechanism is the presence of the triplet $t_{2g}^5e_g$ metal-centered ($^3$MC) state, which has a Ru-N bond length about 0.4~\AA~larger compared to the ground state \cite{Heully,Alary}. This state is generally considered to play an important role in the decay of the $^{3}$MLCT back to the ground state. There is consensus that the two triplet states are close to each other in energy  \cite{Houten,Allen,Durham,Thompson}. The $^3$MC energy position with respect to the $^{3}$MLCT is experimentally controversial. However, the $^3$MC is likely higher in energy than the  $^{3}$MLCT, considering the stability of the latter, see Fig. \ref{level}. We demonstrate that the $^3$MC effectively mediates the decay from the $^1$MLCT to the $^3$MLCT even in the absence of a direct coupling between the $^1$MLCT and the $^3$MC state. The $^3$MC contribution to the relaxed state is minimal.

\textit{Model and parameters.$-$}  In describing the singlet to triplet MLCT decay, we consider three excited states, $^{1}$MLCT, $^{3}$MLCT,  and $^{3}$MC, labeled as state 1, 2 and 3, respectively. These levels couple to the vibrational/phonon modes of energy $\hbar\omega$ of the surrounding ligands, via the Hamiltonian
\begin{eqnarray}
H_{0}=\sum_{i}E_{i}n_{i}+\hbar\omega
a^{\dagger}a+\lambda_{i}n_{i}\left(a^{\dagger}+a\right),\label{h0}
\end{eqnarray}
where $n_{i}$ gives the occupation of state $i$ and $a^{\dagger}$ is the creation operator for the vibrational mode. The coupling is dominated by the stretching mode that changes the metal-ligand bond length. Variations in coupling strength $\lambda_{i}$ lead to different metal-ligand equilibrium distances for the different states. The electron-phonon self-energy is $\varepsilon_{i}=\lambda_{i}^{2}/(\hbar\omega)$.  Since a translation of the coordinates shifts all the $\lambda_{i}$ by a constant in $H_0$, only the relative change in coupling is of importance.  Therefore, we define the electron-phonon self-energy difference $\varepsilon_{ij}=\left(\lambda_{i}-\lambda_{j}\right)^{2}/(\hbar\omega)$ between state $i$ and $j$. In addition, the energy difference between the states after diagonalization of $H_0$ is defined as $\Delta_{ij}=\left(E_{i}-\varepsilon_{i}\right)-\left(E_{j}-\varepsilon_{j} \right)$. The different states couple to each other with  an interaction of strength $V_{ij}$, 
\begin{eqnarray}
H_{1} & = & \sum_{ij}V_{ij}(c_{i}^{\dagger}c_{j}+{\rm H.c.}),
\end{eqnarray}
 where $c_{i}^{\dagger}c_{j}$ causes a particle-conserving transition between states $j$ and $i$. The size and nature of the coupling constants $V_{ij}$ is discussed below. The system is not separated from its surrounding. Exchange of phonons with the bath damps the metal-ligand bond length oscillations via intramolecular energy redistribution processes. The vibrational relaxation by the bath is taken into account by the use of a dissipative Schr\"{o}dinger equation \cite{Veenendaal,chang}.

For a quantitative description, we need to determine the parameters in a reasonable range based on both experiment and theory. Fluorescence spectra show $^{1}$MLCT emission between 500 and 575 nm and  $^{3}$MLCT features around 620 nm \cite{Cannizzo}. This corresponds to an energy difference $\Delta_{12}=0.3$ eV. The electron-phonon self-energy difference $\varepsilon \propto (\Delta x)^2$ is related to $\Delta x$, the difference of the metal-ligand equilibrium distances between two different states. Since the singlet and triplet MLCT have almost the same metal-ligand equilibrium distances \cite{Gawelda}, $\varepsilon_{12}\cong 0.0$ eV. The difference in metal-ligand bond length between the metal-centered and  MLCT  states  is several tenths of an~\AA, corresponding to a change of electron-phonon self-energy of the order of several tenths of an eV  \cite{Bressler,Ordejon} and we take $\varepsilon_{13}=\varepsilon_{23}=0.4$ eV. For the Ru-ligand stretching mode, we take a typical energy  of $\hbar\omega=$ 30 meV \cite{Tuchagues}. The coupling between the singlet and triplet MLCT states is the  spin-orbit coupling of ruthenium, $V_{12}=0.12$ eV \cite{Khudyakov}. There is no direct coupling between the $^1$MLCT and $^3$MC.  In pure octahedral symmetry, there is also no coupling between the $^3$MLCT and the $^3$MC, since the $e_g$ states would not couple to the $\pi^*$ states. However, since most Ru-complexes have a lower symmetry, a small hybridization between the two triplet states should be present. We take a weak coupling constant of $V_{23}=0.08$ eV, where the actual value will depend on the amount of distortion. The environmental relaxation time for a single phonon mode is taken as 40 fs.
\begin{figure}[t]
 \includegraphics[width=1.00\columnwidth]{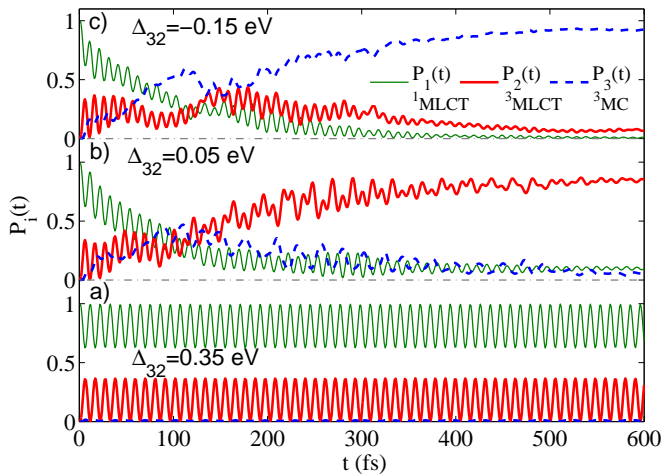}

\caption{(color online) The probability of the $^1$MLCT ($P_1(t)$, green thin solid), the $^3$MLCT ($P_2(t)$,  red solid), and the $^3$MC ($P_3(t)$, blue dashed) as a function of  time for different values of the energy difference $\Delta_{32}$ between the $^3$MC and  $^3$MLCT states.  (a) $\Delta_{32}=0.35$ eV; (b) $\Delta_{32}=0.05$ eV; (c) $\Delta_{32}=-0.15$ eV.}
\label{timee} 
\end{figure}

\textit{Rate Equations.$-$} First, we demonstrate that the decay from singlet to triplet MLCT states cannot be understood from phenomenological rate equations based on the Franck-Condon factors \cite{Hauser,Forster,chang}. The relaxation constant from state 1 ($^1$MLCT) to state 2 ($^3$MLCT) is
$\Gamma=2\pi F_{n}V_{12}^{2}/\hbar^{2}\omega$  where $V_{12}$ is the spin-orbit coupling between singlet and triplet levels and $F_{n}=e^{-g}g^{n}/n!$ is the Franck-Condon factor where $n\approx\Delta_{12}/\hbar\omega$ and $g=\varepsilon_{12}/\hbar\omega$ is the Huang-Rhys factor. In the spin crossover in divalent Fe-complexes, electrons are transferred from $t_{2g}$ to $e_g$ orbitals, resulting in a change of bond length on the order of tenths of an~\AA. The corresponding $\varepsilon$ is tenths of an eV. However, in Ru complexes, the Ru-N bond length changes only about 0.03~\AA, i.e. an order of magnitude less \cite{Gawelda}.  Correspondingly, the change in phonon self-energy $\varepsilon_{12}\sim (\Delta x)^2$ should be two orders of magnitude smaller than in typical Fe-complexes. Since this is negligible with respect to $\Delta_{12}$, it is impossible to meet the condition for ultra-fast decay that the energy gap is near the electron-phonon selfenergy difference \cite{chang}. The lifetime is therefore expected to be much longer than picoseconds (ps). Since there is also no direct coupling between state 1 and 3 ($^3$MC), the initial state 1 cannot reach state 2 either directly or via state 3. Therefore, phenomenological rate equations cannot provide a satisfactory explanation of the singlet to triplet MLCT ultrafast decay.

\textit{Singlet-to-triplet MLCT decay$-$}  In the following we investigate the time-dependence of the states based on quantum-mechanical principles  \cite{Veenendaal,chang}. First, in the absence of state 3 ($^3$MC), we find, as expected,  no decay but an oscillatory behavior involving the triplet and singlet MLCT states as a result of the spin-orbit coupling (not shown). When the $^3$MC is above state 1 ($^1$MLCT), the situation is very similar and no decay is observed, see Fig. \ref{timee}(a) for $\Delta_{32}=0.35$ eV ($\Delta_{31}=0.05$). The population of the $^3$MC state is almost negligible. There is no decay and an an oscillating population of the singlet and triplet MLCT states. The oscillation period of 11 fs is close to frequency expected from the  energy gap $\sqrt{\Delta_{12}^2+4 V_{12}^2}$ in the absence of the $^3$MC state. 

Let us now consider the situation where the $^3$MC state is located between the MLCT states, see Fig. \ref{timee}(b) with $\Delta_{32}=0.05$ eV ($\Delta_{31}=-0.25$ eV).
Ultrafast decay occurs, and after about 300 fs the average populations of the system become stable. The population of the $^3$MLCT state increases to about 85\% at 600 fs. Although, at first sight, the time dependence of the probabilities resembles a cascading from state 1 ($^1$MLCT) to 2 ($^3$MLCT) via state 3 ($^3$MC), it is important to remember that in the model there is no direct coupling  between state 1 and 3 ($V_{13}=0$). Therefore, in a classical rate equation, the expected decay time would be very large, $T_{13}=(2\pi F_{n}V_{13}^{2}/\hbar^{2}\omega)^{-1}\rightarrow \infty$.  A detailed analysis of the  evolution of the state occupations with time, shows that the initial $^1$MLCT
state couples to the lowest vibrational level of the $^3$MLCT via the spin-orbit coupling. Minor distortions from octahedral symmetry allow a hopping $V_{23}$ of the electron from the ligand $\pi^*$ orbitals into the ruthenium $e_g$ orbital, populating the $^3$MC state. However, due to the large change in electron-phonon self-energy $\varepsilon_{23}$, the hopping to the lowest vibrational $^3$MC state  is strongly reduced due to the Franck-Condon factors. Therefore, the hopping between the $^3$MLCT and $^3$MC does not conserve phonons. This allows the system to decrease its energy via  vibrational cooling due to intramolecular energy redistributions \cite{Veenendaal}. The probability in the $^3$MC state can however also couple back again to the $^3$MLCT, again leading to the creation of excited phonon states that are subject to  vibrational cooling. Therefore, the interplay between the $^3$MLCT and the  $^3$MC state leading to the creation of excited phonons allows the local system to dissipate energy via the damping of the Ru-ligand bond-length oscillations. The consequence is a reduction of the recurrence of the $^1$MLCT state. In the relaxed state ($\gtrsim 500$ fs), the contributions of the $^3$MC is small after reaching a maximum of close to 50\% around 100-150 fs. The contribution of the $^1$MLCT is of the order of 10\%, justifying the traditional designation of the final state of the decay in Ru-complexes as a triplet. 

In addition, let us consider the case that the $^3$MC state is lower in energy than the $^3$MLCT state. For $\Delta_{32}=-0.15$ eV, we observe a cascading process from the $^1$MLCT via the $^3$MLCT to the $^3$MC state, see Fig. \ref{timee}(c). The $^3$MC  reaches an occupation of approximately 92\% at about 600 fs. In a real system, the occupation of the $^3$MC state  most likely decays back to the ground state (not included in the calculation).

\begin{figure}[t]
\includegraphics[width=1.0\columnwidth]{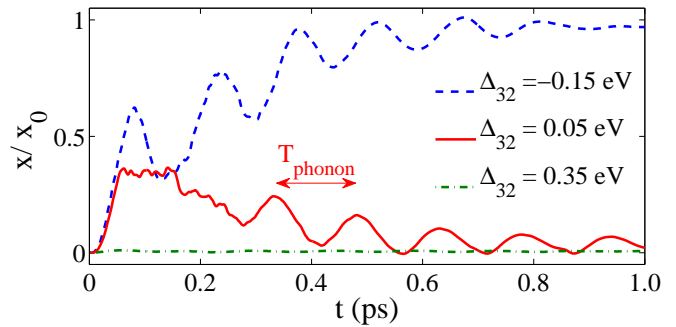}
\caption{(color online) The metal-ligand bond length as a function of time $t$ for different energy differences $\Delta_{32}$ between the $^3$MC and $^3$MLCT states. The bond length change is normalized to the
bond elongation ($x_0=0.4$~\AA) of the MC triplet state with respect to the MLCT state  The oscillation period is close to the phonon's $T_{\rm phonon}=$137.8 fs.}
\label{ax} 
\end{figure}
Finally, we study the evolution of the metal-ligand bond length as a function of time,  see Fig. \ref{ax}. EXAFS measurements \cite{Bressler} taken 70 ps after the photo-excitation show a bond length contraction of about 0.03~\AA~with respect to measurements in the absence of a photoexcitation. This rules out a strong contribution of the $^3$MC, which is expected to have a significantly larger bond length due to the presence of electron density in the $e_g$ orbitals. The small change in bond length is supported by our model. 
 The calculated bond length elongation is strongly related to the probability of state 3, see Fig. \ref{timee}. For the situation, where the $^3$MC state is located between the MLCT state ($\Delta_{32}=0.05$ eV), the bond length initially increases  and reaches a maximum of about 0.16~\AA~ around 100 fs. Subsequently, the length decreases as the occupation of the $^3$MLCT state increases. At about 600 fs, the average bond elongation is only 0.024~\AA~larger with respect to the MLCT states. This should be contrasted with the situation where the  $^3$MC state is lower in energy than the $^3$MLCT ($\Delta_{32}=-0.15$ eV). Here, the bond length increases by several tenths of an~\AA, as the $^3$MC is occupied. On the other hand, if the $^3$MC state energy is higher than that of the $^1$MLCT ($\Delta_{32}=0.35$ eV), the bond length is also almost the same as that of the MLCT states.

\textit{Conclusion.$-$}  
Experiments on ruthenates over several decades have established that the photo-excited singlet metal-to-ligand state decays into its triplet counterpart with very little change in lattice parameter \cite{Demas,Damrauer,Gawelda}. This contradicts the expectation that, in the absence of a clear decay mechanism, an oscillation should occur between the singlet and triplet MLCT states as a result of the strong ruthenium spin-orbit coupling. 
This paper has shown that a triplet metal-centered state located between the MLCT states can effectively mediate the decay from the MLCT singlet to the triplet state. The decay cannot be simply described in terms of a phenomenological rate equation, but is a result of vibrational cooling of the oscillations of the Ru-ligand bond length due to the intersystem crossings between the $^3$MLCT and the $^3$MC states. The presence of the $^3$MC also strongly reduces the mixing of singlet and triplet states justifying the standard singlet and triplet notation which at first appears erroneous for a compound with a large spin-orbit coupling. The decay occurs on a timescale of 300-400 fs and has a very high quantum  efficiency.

Several aspects provide interesting experimental tests of the described mechanism. First, although the metastable state reached after several hundreds of femtoseconds shows little change in the metal-ligand bond length, there is an increase in bond length of the order of 0.15-0.2~\AA~in the 200-300 fs following the photoexcitation. The observation of this dynamic response is in the reach of state-of-the-art optical pump/x-ray probe techniques \cite{Bressler}. Second, the coupling between the $^3$MLCT and $^3$MC states is a result of finite hopping matrix elements due to distortions that break the octahedral symmetry. The dependence of the decay time in different geometries can provide additional information on the detailed nature of the decay mechanism.

{\it Acknowledgments}.$-$ We are thankful to Yang Ding for helpful discussions. This
work was supported by the U.S. Department of Energy (DOE), DE-FG02-03ER46097, and NIU's Institute for Nanoscience, Engineering, and Technology. Work at Argonne National Laboratory was supported by the U.S. DOE,
Office of Science, Office of Basic Energy Sciences, under contract DE-AC02-06CH11357.

\end{document}